\documentclass[prx,twocolumn,english,superscriptaddress,floatfix,longbibliography]{revtex4-2} 
\usepackage{bm}         
\usepackage{verbatim}
\usepackage{multirow}
\usepackage{xcolor} 
\usepackage{amsmath}
\usepackage{amsfonts}
\usepackage{soul}
\usepackage{xr-hyper}
\usepackage{makecell}
\usepackage{titlesec}
\usepackage{ragged2e}
\usepackage[utf8]{inputenc}
\usepackage{amsmath,amsfonts,amssymb}
\usepackage{graphicx}
\usepackage{mathtools}
\usepackage[colorlinks]{hyperref}
\usepackage{fancyhdr}
\usepackage[normalem]{ulem}
\usepackage{float}
\usepackage{placeins}

\hypersetup{colorlinks=true, pdfstartview=FitV, linkcolor=blue, citecolor=black, plainpages=false, pdfpagelabels=true, urlcolor=blue}

\restylefloat{table}
\usepackage[normalem]{ulem}
\hypersetup{colorlinks=true, pdfstartview=FitV, linkcolor=blue, citecolor=black, plainpages=false, pdfpagelabels=true, urlcolor=blue}

\begin{document}

\newcommand{\ahat}{\hat{a}}
\newcommand{\ahatd}{\hat{a}^\dagger}
\newcommand{\rhat}{\hat{r}}
\newcommand{\rhatd}{\hat{r}^\dagger}

\title{Entangling interactions between artificial atoms mediated by a multimode left-handed superconducting ring resonator}
\author{T. McBroom-Carroll}
\affiliation{Department of Physics, Syracuse University, Syracuse, New York 13244-1130 USA}

\author{A. Schlabes}
\affiliation{Peter Grünberg Institute (PGI-2), Forschungszentrum Jülich, Jülich 52428, Germany}
\affiliation{Institute for Quantum Information, RWTH Aachen University, D-52056 Aachen, Germany}
\affiliation{Jülich-Aachen Research Alliance (JARA), Fundamentals of Future Information Technologies, Germany}

\author{X. Xu}
\affiliation{Peter Grünberg Institute (PGI-2), Forschungszentrum Jülich, Jülich 52428, Germany}
\affiliation{Jülich-Aachen Research Alliance (JARA), Fundamentals of Future Information Technologies, Germany}

\author{J. Ku}
\affiliation{Department of Physics, Syracuse University, Syracuse, New York 13244-1130 USA}

\author{B. Cole}
\affiliation{Department of Physics, Syracuse University, Syracuse, New York 13244-1130 USA}

\author{S. Indrajeet}
\affiliation{Department of Physics, Syracuse University, Syracuse, New York 13244-1130 USA}

\author{M. D.  LaHaye}
\affiliation{Air Force Research Laboratory, Information Directorate, Rome NY 13441 USA}
\affiliation{Department of Physics, Syracuse University, Syracuse, New York 13244-1130 USA}

\author{M. H. Ansari}
\affiliation{Peter Grünberg Institute (PGI-2), Forschungszentrum Jülich, Jülich 52428, Germany}
\affiliation{Institute for Quantum Information, RWTH Aachen University, D-52056 Aachen, Germany}
\affiliation{Jülich-Aachen Research Alliance (JARA), Fundamentals of Future Information Technologies, Germany}

\author{B. L. T. Plourde}
\email[]{bplourde@syr.edu}
\affiliation{Department of Physics, Syracuse University, Syracuse, New York 13244-1130 USA}
\begin{abstract}
Superconducting metamaterial transmission lines implemented with lumped circuit elements can exhibit left-handed dispersion, where the group and phase velocity have opposite sign, in a frequency range relevant for superconducting artificial atoms. Forming such a metamaterial transmission line into a ring and coupling it to qubits at different points around the ring results in a multimode bus resonator with a compact footprint. Using flux-tunable qubits, we characterize and theoretically model the variation in the coupling strength between the two qubits and each of the ring resonator modes. Although the qubits have negligible direct coupling between them, their interactions with the multimode ring resonator result in both a transverse exchange coupling and a higher order $ZZ$ interaction between the qubits. As we vary the detuning between the qubits and their frequency relative to the ring resonator modes, we observe significant variations in both of these inter-qubit interactions, including zero crossings and changes of sign. The ability to modulate interaction terms such as the $ZZ$ scale between zero and large values for small changes in qubit frequency provides a promising pathway for implementing entangling gates in a system capable of hosting many qubits.
\end{abstract}

\maketitle
\section{Introduction}
In the field of circuit quantum electrodynamics (cQED), linear resonators are commonly used for readout of artificial atom qubits, and in some cases, for coupling between qubits \cite{blais2021circuit,blais2004cavity}. Variants on this architecture are the dominant paradigm for current superconductor-based quantum processors. The resonators are often formed from planar transmission lines with a single mode near the frequency range of the qubits, resulting in physically large footprints of several mm. Utilizing an even larger physical footprint, devices with dense mode spectra near the qubit frequency range have been realized using ultra-long linear resonators \cite{sundaresan2015beyond}. Similar multimode cQED systems have been studied for implementing quantum memories \cite{naik2017random} and quantum simulations \cite{cirac2012goals,armstrong2012programmable}.

The use of metamaterials formed from lumped-element inductors and capacitors allows for the implementation of transmission lines with unconventional wave dispersion. In the case of left-handed dispersion \cite{veselago2002electrodynamics}, the wave frequency is a falling function of wavenumber above an infrared cutoff frequency $f_{IR}$, below which waves are unable to propagate \cite{eleftheriades2002planar,caloz2004transmission}. In the context of cQED, left-handed metamaterials produce a dense spectrum of orthogonal microwave modes above $f_{IR}$, which can be engineered to fall in the frequency range of conventional superconducting qubits \cite{wang2019mode,indrajeet2020coupling}.

Ring resonators have been used in integrated photonics systems to form compact optical resonances, or whispering gallery modes, for a broad range of applications including microwave-optical transducers, microwave frequency combs, and multimode nonlinear optics \cite{shao2019microwave, singh2019quantum, ren2016highly, kippenberg2018dissipative, moody20222022}. Superconducting ring resonators with right-handed dispersion have been used in cQED applications resulting in novel properties \cite{huang2021superconducting,hazra2021ring}, but require a large footprint to ensure a minimum of one wavelength matches the circumference.

In this manuscript, we describe a superconducting ring resonator formed from a left-handed metamaterial transmission line with two transmon qubits coupled at different points around the ring. The left-handed dispersion results in a dense spectrum of modes in the vicinity of the transition frequencies for the qubits. By wrapping the left-handed transmission line into a ring structure, we produce a device with a relatively compact footprint with unique properties that arise due to the symmetry of the ring. We perform a detailed modeling of the standing-wave structure and degeneracy breaking in the ring resonator, allowing us to predict the coupling energy scales between the qubits and each ring resonator mode \cite{Ansari2019beyondDispersion}. The multimode coupling between the qubits with the ring resonator serving as a bus results in significant variations in both the transverse exchange coupling between the qubits as well as higher order $ZZ$ interactions as the qubits are tuned between different frequency regimes. 

Our theoretical modeling of these interactions is in close agreement with our experimental measurements. The ability to modulate the $ZZ$ interaction strength between zero and a large value for a small change in qubit frequency allows for the possibility of implementing fast, high-fidelity entangling gates between pairs of qubits located around the ring resonator.

\begin{figure}[h!]
    \centering
    \includegraphics{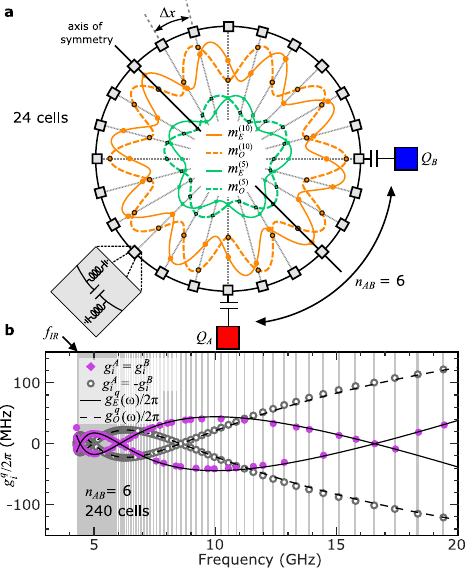}
    \caption{(a) Mode structure for two pairs of degenerate ring resonator modes. The ring resonator consists of 24~unit cells, shown with gray squares, with each cell containing a capacitor shunted to ground by two inductors. One pair of even (odd) modes with ${k\Delta x=10\tfrac{2\pi}{N}}$ is shown in solid (dashed) orange lines, labeled $m^{(10)}_{E}$($m^{(10)}_{O}$); a second pair with ${k\Delta x=5\tfrac{2\pi}{N}}$ is depicted with solid (dashed) green lines, labeled $m^{(5)}_{E}$($m^{(5)}_{O}$). The coupling strength of $Q_A$ and $Q_B$ to the mode is determined by the amplitude of the wave at the cell where each qubit is capacitively coupled to the resonator. (b) Theoretical $g$-coupling values for $Q_A$ and $Q_B$, coupled to a hypothetical 240-cell ring resonator with qubit separation $n_{AB}$ = 6. Simulated data points at which the coupling for $Q_A$, $g_i^A/2\pi$, and for $Q_B$, $g_i^B/2\pi$, share the same sign for a given mode are shown in purple and follow Eq.~(\ref{eq:gE}). Points at which the $g_i^A$ and $g_i^B$ coupling to $Q_A$ and $Q_B$, respectively, have opposite signs are shown in gray and are associated with Eq.~(\ref{eq:gO}).}
    \label{fig:theory}
\end{figure}

\section{Ring resonator modes and qubit coupling: theory}
\noindent A resonator formed from a conventional transmission line (TL) can be modeled as an array of $LC$ unit cells, each with inductors shunted by capacitors to ground \cite{pozar2011microwave}. 
For this discrete model, a linear TL of length $d_l$ is partitioned into $N$ unit cells, each with size $\Delta x=d_l/N$. The cell capacitance (inductance) are  $C_L=c_L  \Delta x$ ($L_L=l_L \Delta x$) with $c_L$ ($l_L$) being the respective quantities per unit length. In the continuum limit $N\to \infty$, the lumped-element model reproduces the linear dispersion of transverse electromagnetic modes $\omega(k) = k/\sqrt{c_L l_L}$, with $k$ being wave number. This configuration results in a right-handed resonator, as its wave, electric, and magnetic vectors make a right-handed set in three dimensions. However, switching the positions of the capacitors and inductors results in a left-handed dispersion relation \cite{eleftheriades2002planar}, with $\omega$ decreasing for increasing $k$:
\pagestyle{fancy}
\fancyfoot[C]{ \small Approved for Public Release; Distribution Unlimited. PA \#: AFRL-2024-0930.}
\fancyhead[R]{\thepage}
\begin{equation}\label{eq:dispersion}
\omega(k)=\frac{1}{2\sqrt{L_L C_L}}\frac{1}{\sin(|k| \Delta x/2)}.
\end{equation}
In an ideal left-handed metamaterial, varying wave number in the domain of $|k\Delta x| \leq \pi$ shows no upper bound on frequency in the $k\rightarrow 0$ limit. 
In actual devices, a finite upper bound results because these modes hybridize with parasitic right-handed modes. Additionally, left-handed metamaterials exhibit a lower bound at $|k\Delta x | \to  \pi$, referred to as the infrared cutoff frequency $\omega_{IR} = 1/{2\sqrt{C_LL_L}}$, which is only slightly perturbed with parasitic modes \cite{lai2004composite}. In between the upper and lower frequency bounds there are a finite number of modes, with those close to the infrared cutoff being densely grouped, and mode separation increasing monotonically  with higher frequencies \cite{wang2019mode}. \par

Next, we consider a left-handed ring resonator design made by connecting the two ends of a left-handed TL. This enforces periodic boundary conditions, similar to Born-von Karman boundary conditions in solid state physics \cite{ashcroft2022solid}. This condition highlights two types of traveling waves in the ring resonator: clockwise and counterclockwise moving waves. Given that the dispersion relation in Eq.~(\ref{eq:dispersion}) depends on $|k|$, waves moving in opposite directions have the same frequency. In a perfect ring with identical cells and no external coupling, equal combinations of clockwise $|k\rangle$ and counterclockwise $|-k\rangle$ waves define the standing wave of frequency $\omega(k)$, with even and odd parity $|E\rangle=(|k\rangle + |-k\rangle)/\sqrt{2}$ and  $|O\rangle=(|k\rangle - |-k\rangle)/\sqrt{2}$, respectively. The parity is indistinguishable at $k=0$. Similarly, for the mode at the IR cutoff, the difference between the clockwise and counterclockwise argument $k\Delta x$ is $2\pi$, thus the two parities are indistinguishable. This is akin to states at the edge of the Brillouin zone in solid state systems. A total of $N/2-1$ degenerate parity pairs can be found in an ideal ring resonator. Figure~\ref{fig:theory}(a) shows two pairs of such degenerate modes in the form of standing waves with a relative phase shift of $\pi/2$. In the subsequent sections, we will show that any source of asymmetry between cells or local external coupling to the ring lifts the degeneracy. \par
Let us consider a ring resonator that is capacitively coupled to two weakly-anharmonic qubits, $Q_A$ and $Q_B$, separated by $n_{AB}$ cells around the ring [Fig.~\ref{fig:theory}(a)]. Using the standard circuit quantization within the rotating wave approximation, the total Hamiltonian reads \cite{blais2007quantum}
\begin{align}
\label{eq:H}
H/\hbar=& \sum_{q} \Big[\omega_q \ahatd_q\ahat_q + \frac{\delta_q}{2} \ahatd_q\ahat_q (\ahatd_q\ahat_q-1)\Big]  \nonumber\\
 + &\sum_{|k|,P}\omega(k)\rhatd_{k,P}\rhat_{k,P} + \sum_{|k|,P,q}g^q_{P}(k)(\rhatd_{k,P}\ahat_q + \rhat_{k,P}\ahatd_q),
\end{align}
with  $\hat{a}_q^\dagger$($\hat{a}_q$) being the creation(annihilation) operator of qubit $q=A,B$ and $\delta_q$ being the qubit anharmonicity. Here $\hat{r}_{k,P}^\dagger$($\hat{r}_{k,P}$) are the creation(annihilation) operator of ring resonator mode $k$ and parity $P=E,O$, and the qubit-resonator mode coupling strengths $g_{P}^q(k)$ are each proportional to the amplitude of the standing wave for the particular mode at the qubit location, as shown in Fig.~\ref{fig:theory}(a), and therefore depend on $k$ and parity $P$. The even- and odd-parity coupling strengths can be expressed as:
\begin{align}
\label{eq:gE}
g_{E}^{B}(k)=g_{E}^{A}(k) \propto &\frac{\omega^\frac{3}{2}(k)}{\sqrt{N}}\cos \left(k\frac{x_{AB}}{2}\right) \\
\label{eq:gO}
-g_{O}^{B}(k)= g_{O}^{A}(k) \propto  & \frac{\omega^\frac{3}{2}(k)}{\sqrt{N}}\sin \left(k\frac{x_{AB}}{2}\right),
\end{align}
with $x_{AB}= n_{AB} \Delta x$ being the distance between the qubits (\ref{sec:gdetails}). 
Here, the justification for the terms `even' and `odd' parity becomes clear, as even modes couple with the same sign to qubits $Q_A$ and $Q_B$, while odd modes couple with opposite signs to the two qubits. The coupling strengths depend periodically on the product of $k$ and $x_{AB}$.  In Fig.~\ref{fig:theory}(a), the solid line is a cosine and the dashed line is a sine mode. The origin is located in the middle between the qubits, thus clarifying the signs in Eqs.~(\ref{eq:gE},\ref{eq:gO}), as $Q_A$ is connected to cell number $n_{AB}/2$,  and $Q_B$ to $-n_{AB}/2$. 

Figure \ref{fig:theory}(b) shows the oscillatory nature of the $g$-couplings at different mode frequencies. For better visualization of the oscillations, we consider a ring with $N=240$ unit cells, but keep the qubit-qubit cell separation $n_{AB}=6$. Since the cosine and sine functions do not depend on $N$, this does not affect the oscillations in coupling strength. We compare the analytical expressions of Eqs.~(\ref{eq:gE},\ref{eq:gO}) with the numerical simulations that predict the same odd (gray) and even (purple) parity behavior. Note that the parity of the $g$-couplings is crucial for calculation of the effective $J$-coupling between qubits presented later in the manuscript.

\begin{figure}[h!]
    \centering \includegraphics{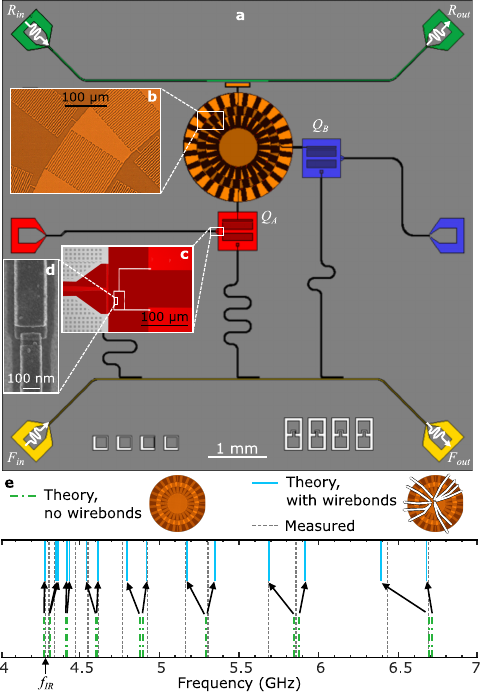}
    \caption{(a) Chip layout for metamaterial ring-resonator device. (b, c) Optical micrographs of device with false-color highlighting. (d) SEM image of Josephson junction in qubit. Device has two flux-tunable transmon qubits, $Q_A$ and $Q_B$, coupled to a ring resonator at the positions indicated. $R_{in}$/$R_{out}$ connections to upper feedline used to probe ring resonator modes. $F_{in}$/$F_{out}$ connections used for measuring readout resonators coupled to each qubit. (e) Measured ring resonator mode frequencies are shown with gray dashed lines. Theoretical mode frequencies are shown with green dashed-dotted lines when stray inductance due to wirebonds is not included and any degeneracy lifting is due to the qubits or feedline. Solid blue lines show large degeneracy lifting effect of wirebonds on the ring resonator mode frequencies.}
    \label{fig:chip}
\end{figure}

\section{Device Layout}
\noindent The device consists of a 24-cell, left-handed superconducting metamaterial ring-resonator comprised of interdigitated capacitors with double-sided meander-line inductors to ground (Fig.~\ref{fig:chip}).
The ring resonator is capacitively coupled to a feedline ($R_{in}/R_{out}$) [Fig.~\ref{fig:chip}(a)] for probing ring-resonator modes. Two flux-tunable, floating-style asymmetric transmon qubits \cite{hutchings2017tunable}, $Q_A$ and $Q_B$, are capacitively coupled to the ring resonator 
90 and 180 degrees from the feedline coupling point, respectively. Both qubits have readout resonators inductively coupled to a separate feedline ($F_{in}/F_{out}$), as well as separate on-chip flux-bias lines for tuning the transition frequency (\ref{sec:Fabrication} and \ref{sec:Device}). 

Figure~\ref{fig:chip}(e) shows the location of aluminum wirebonds used to connect the central superconducting disk inside the ring resonator to the ground plane of the chip. These ground bonds have non-negligible inductance, which we estimate to be 1~nH/mm \cite{wenner2011wirebond}. The bonds are not spaced symmetrically due to the adjacent locations of the qubits and feedline. These wirebonds break the symmetry of the clockwise and counterclockwise propagating waves and lift the degeneracy of the even and odd ring-resonator modes. The infrared cutoff frequency for this ring resonator device is 4.287~GHz, as shown in Figure~\ref{fig:chip}(e), below which we observe no modes. Between 4-6.5~GHz, we measure thirteen modes to which the qubits may couple. As expected, the modes above the infrared cutoff come in pairs that have broken degeneracy due to the presence of the qubits, the feedline, and the wirebonds. The size of the degeneracy lifting ranges from $\sim$30~MHz near the IR mode to $\sim$250~MHz at higher frequencies and is in good agreement with theoretical simulations including a model for the wirebonds, as shown in Fig.~\ref{fig:chip}(e) in blue. The same simulation without wirebonds fails to reproduce the observed lifting of degeneracies 
(\ref{sec:wirebond}).

\begin{figure}
    \centering
    \includegraphics{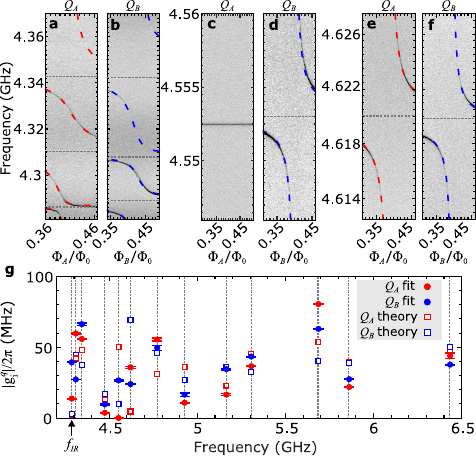}
    \caption{Vacuum Rabi splittings measured via $R_{in}$/$R_{out}$, between three ring resonator modes and (a) $Q_A$, (b) $Q_B$. Dashed lines in red (blue) show fits to splitting data for $Q_A$ ($Q_B$) obtained from reduced Hamiltonian including one qubit and the three modes (see \ref{sec:gexp}). Horizontal dotted lines show dressed mode frequencies. (c-f) Splittings and reduced Hamiltonian fits for two single modes for each qubit. For all measurements, the spectator qubit is fixed at its upper flux insensitive sweet spot ($\Phi_q/\Phi_0$ = 0). As shown in (c), $Q_A$ has effectively zero coupling to the ring resonator mode at 4.553~GHz, while (d) shows $Q_B$ has a coupling strength of 26.6~MHz. (g) Extracted and theoretical $|g_i^q|/2\pi$ values 
    for each ring-resonator mode $i$.  
    Error bars on experimental data  computed from 95\% confidence intervals in $g$-coupling fits (see also \ref{sec:gexp}). Vertical dashed lines show bare ring resonator mode frequencies.}
    \label{fig:gcoupling}
\end{figure}
%
\section{Mode structure and {\it g}-coupling}
\noindent To characterize the coupling strength between each qubit and the different ring resonator modes, we probe the modes by measuring the microwave transmission through the ring resonator feedline ($R_{in}/R_{out}$) while scanning the flux for one of the qubits with the other qubit flux-biased at its upper flux-insensitive sweetspot (\ref{sec:Fridgewiring} includes details on cabling and readout). We perform these measurements for each of the thirteen modes between 4-6.5~GHz, and observe vacuum Rabi splittings when each qubit hybridizes with the ring-resonator mode photonic state. A selection of these measurements for $Q_A$ and $Q_B$ is shown in Fig.~\ref{fig:gcoupling}[a-f]. To extract the magnitude of $g_i^{A}$($g_i^{B}$) for $Q_A$($Q_B$), we perform a least-squares minimization to fit each set of splitting data to a reduced Hamiltonian for the ring-resonator mode-qubit system, shown as dashed lines. In regions where the mode spacing and coupling are comparable, we simultaneously fit multiple modes to extract each $g_i^{A}$ and $g_i^{B}$, as shown in Fig.~\ref{fig:gcoupling}(a,b) (\ref{sec:gexp}). Near $f_{IR}$, the mode spacing is 23~MHz between the IR-cutoff mode and the next mode, while $g_i^A/2\pi$ = 13~MHz and $g_i^B/2\pi=$~36.5~MHz for the IR-cutoff mode. For $Q_A$, one ring-resonator mode near 4.55~GHz [Fig.~\ref{fig:gcoupling}(c)] has effectively zero coupling, while for $Q_B$, the extracted magnitude of $g_i^{B}/2\pi$ is 26.6~MHz to the same mode [Fig.~\ref{fig:gcoupling}(d)]. While theory predicts zero-coupling to the modes at the roots of Eqs.~(\ref{eq:gE},\ref{eq:gO}), this particular mode is not a root. In the simple model without the feedline and wirebonds, this mode has a nonzero coupling strength to $Q_A$; only in the more complex model does this mode result in a negligible coupling strength (\ref{sec:wirebond}).

Experimental and theoretical values for $g_i^{A}$($g_i^{B}$) for $Q_A$($Q_B$) and ring-resonator modes from 4~GHz to 6.5~GHz are shown in Fig.~\ref{fig:gcoupling}(g). Theoretical calculations include a model of the feedline and the wirebonds, the introduction of which requires the use of numerical calculations instead of the analytical expressions in Eqs.~(\ref{eq:gE},\ref{eq:gO}). Despite the added complexity, we can still observe features of the sine and cosine behavior of $g_i^{A/B}$. For example, we find a pair of modes at 4.344~GHz and 4.475~GHz; the former having a large coupling strength, since it is an odd mode and Eq.~(\ref{eq:gO}) has a maximum here, while the latter is even and has negligible coupling, following Eq.~(\ref{eq:gE}).


We observe a broad array of coupling magnitudes ranging from 0-80~MHz. While the theoretical model captures the general behavior of $g_i^{A/B}$, a more complete model of the symmetry-breaking perturbations of the ring-resonator circuit is required for better quantitative agreement. Although we cannot directly measure the sign of $g$, the magnitude and parity of the pairs of $g_i^{A}$ and $g_i^{B}$ for $Q_A$ and $Q_B$ to each mode $i$ impacts the qubit-qubit interactions that are mediated by the ring resonator. We utilize these extracted $g$-couplings for each of the modes in the subsequent sections for analyzing our measured qubit-qubit interactions.

\vspace{2mm}
\section{{\it J}-coupling between qubits}

\begin{figure}[h!]
    \centering
    \includegraphics{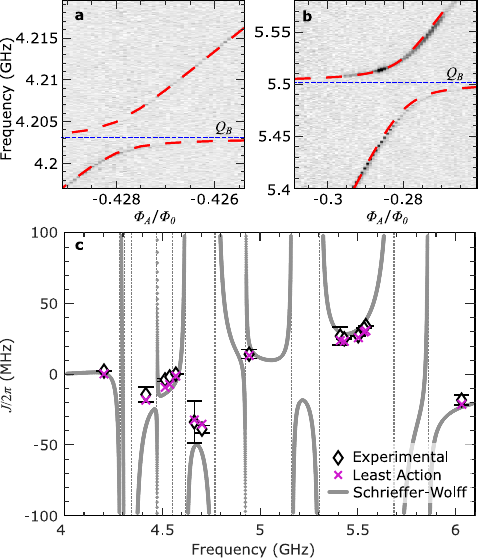}
    \caption{(a) and (b) show qubit spectroscopy for $Q_A$ as a function of flux when $Q_B$ is at a fixed frequency, denoted by a blue dotted line. Red dashed lines show fits to data obtained via minimization of reduced Hamiltonian from which we determine $J/2\pi$ (see \ref{sec:Jdetails}). (c) Experimental and theoretical $J$-coupling values as a function of frequency. Diamonds show experimental values for $J/2\pi$, with error bars computed from 95\% confidence intervals in fits (see also \ref{sec:Jdetails}). Gray lines and purple crosses show theoretical $J/2\pi$ values calculated using Schrieffer-Wolff and Least Action, respectively. Dotted vertical lines show bare ring-resonator mode frequencies.} 
    \label{fig:Jcoupling}
\end{figure}

\noindent We next explore the transverse exchange coupling between the qubits mediated by virtual photon exchange \cite{majer2007coupling} with the various ring-resonator modes. By using the qubit readout resonators to perform conventional qubit spectroscopy and the local flux-biasing capability to independently tune the qubits, we fix the frequency of one qubit, say $Q_B$, while flux-tuning the other qubit ($Q_A$). A nonzero $J$-coupling between the qubits results in an anti-crossing in the spectrum when the bare qubit frequencies cross. We can adjust the frequency of the crossing point for the bare qubit frequencies relative to the ring-resonator modes and study the variation of the exchange coupling between the qubits (Fig. \ref{fig:Jcoupling}).

To extract the magnitude of $J$, we fit the spectroscopy data with a reduced Hamiltonian model in which the qubits are restricted to two states and only the nearest ring-resonator modes are accounted for (\ref{sec:Jdetails}). In the spectroscopy measurements in Fig.~\ref{fig:Jcoupling}, we observe a small exchange coupling of 2.1~MHz when the qubits are biased below $f_{IR}$ [Fig.~\ref{fig:Jcoupling}(a)], while the $J$-coupling is more than an order of magnitude larger when the qubits are biased in the dense portion of the ring-resonator mode spectrum [Fig.~\ref{fig:Jcoupling}(b)].

Figure~\ref{fig:Jcoupling}(c) shows theoretical predictions for $J$ compared to extracted values for a range of qubit frequencies. In the perturbative Schrieffer-Wolff approximation, $J$ is given by a sum over all modes
\begin{equation}
    J = \frac{1}{2}\sum_{i=1}^N g^A_ig^B_i\left(\frac{1}{\Delta^A_i} + \frac{1}{\Delta^B_i}\right),
\label{eq:J}
\end{equation}
where $\Delta^A_i$($\Delta^B_i$) give the detuning for $Q_A$($Q_B$) to each ring resonator mode $i$. Since $J$ depends on the product $g^{A}_ig^{B}_i$ for each mode, even-parity modes will have a positive contribution to the sum and odd-parity modes will have a negative one, however this can be changed by the sign of the detuning. When the qubits are biased below $f_{IR}$ and all $\Delta_i^q$ have the same sign, $J$ will be small, as shown in Fig. 4(c). 
Although our spectroscopy measurements only provide the magnitude of $J$, using the perturbative theoretical calculations with different possible parities of $g^A_i$ and $g^B_i$ for each mode $i$, we determine the sign of each qubit-mode coupling that provides the best agreement with the exchange coupling measurements (\ref{sec:Jdetails}).
From the experimentally measured $J$ values and the agreement with the theoretical modeling, it is clear that the exchange coupling can vary in magnitude between zero and at least 41~MHz and can change sign depending on the detuning between the qubits and the various ring-resonator modes. Since the mode spectrum is dense and interactions to modes may be strong, perturbation theory is not guaranteed to be a good choice. Therefore we also compare our results to the non-perturbative Least Action method \cite{Xu2021ZZfreedom,magesan2020effective,cederbaum1989block} at selected frequencies. This method agrees well with both experimental and perturbative data, except for cases where the qubit and resonator mode frequency are quite close, in which case the perturbative method diverges (\ref{sec:Jtheory}). 
%

\begin{figure*}
    \centering
    \includegraphics[]{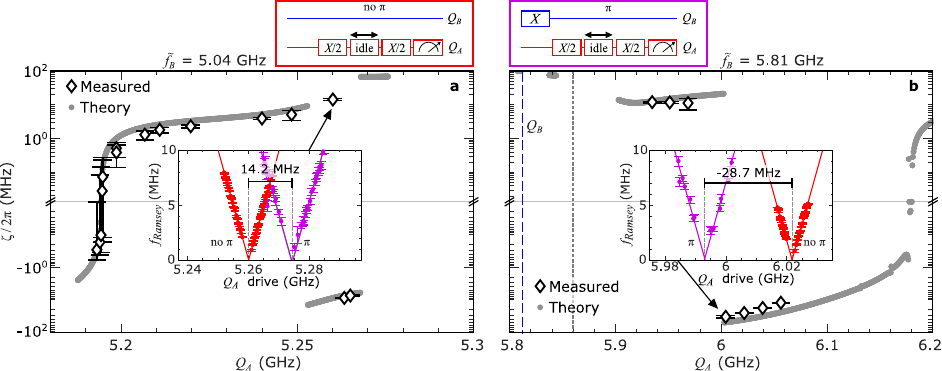}
    \caption{$\zeta/2\pi$ as a function of the dressed frequency of $Q_A$, $\tilde{f_A}$, when the dressed frequency of $Q_B$, $\tilde{f_B}$, is (a) 5.04~GHz, (b) 5.81~GHz (see \ref{sec:ZZdetails}). Measured and theoretical values are shown as diamonds and gray lines, respectively. Vertical dashed lines in blue and gray show the location of $Q_B$ and the ring resonator modes in frequency space, respectively. Insets show Ramsey oscillation frequency fit data as a function of $Q_A$ drive frequency taken from Ramsey fringe measurements.  Red data points and fit lines generated by performing a simple Ramsey pulse sequence (shown in red box) on $Q_A$ at multiple detunings and extracting oscillation frequency. Data resulting from pulse sequence in which a $\pi$-pulse is applied to $Q_B$, followed by a Ramsey on $Q_A$ (shown in purple box). Error bars computed from 95\% confidence intervals for both Ramsey oscillation fits (see also  \ref{sec:ZZdetails}). Intersection of fit lines where Ramsey oscillation frequency vanishes indicate $Q_A$ frequency, from which we compute $\zeta/2\pi$.}
    \label{fig:ZZ}
\end{figure*}
 
%
\vspace{2mm}
\section{{\it ZZ} interactions} 

\noindent In general, interactions between pairs of qubits and resonator modes coupled to the qubits lead to $ZZ$ interactions between the qubits, where the state of one qubit shifts the transition frequency of the other qubit. Such interactions can be problematic in multi-qubit systems and can generate unwanted and uncontrolled entanglement \cite{sheldon2016procedure,magesan2020effective}. Various approaches have been explored for reducing \cite{mundada2019suppression, li2020tunable} or nulling \cite{ku2020suppression,Xu2021ZZfreedom} these interactions. On the other hand, the $ZZ$ coupling can also be used for implementing two-qubit entangling gates, provided the interaction strength can be modulated \cite{Xu2023PFgate,collodo2020implementation,hua2015one,sete2021floating}. 

In the dense multi-mode spectrum of our metamaterial ring resonator system, we explore the $ZZ$ interaction between the two qubits for multiple qubit detunings within different regions of the mode spectrum. We use a standard Ramsey interferometry technique to extract the $ZZ$ interaction strength \cite{takita2017experimental}. First, we bias $Q_B$ to a particular transition frequency, then we adjust the bias of $Q_A$ to various frequencies relative to $Q_B$ and the mode spectrum. At each bias point for $Q_A$, we perform a standard Ramsey fringe sequence while stepping through the drive frequency for the $X/2$ pulses for $Q_A$, thus allowing us to identify the transition frequency for $Q_A$. We then perform a second Ramsey fringe measurement of $Q_A$, but now with each Ramsey sequence preceded by an $X$ pulse on $Q_B$. The difference in the transition frequency for $Q_A$ between the two Ramsey fringe sequences corresponds to the $ZZ$ interaction, $\zeta$ (\ref{sec:ZZdetails}).

In Fig.~\ref{fig:ZZ}(a,b) we show two series of $ZZ$ measurements as a function of the dressed transition frequency for $Q_A$ for two different bias points of $Q_B$. We observe that $\zeta$ can vary over a wide frequency range, covering both positive and negative values and crossing through zero, both smoothly in some regions and discontinuously in others. The theoretical curves are calculated by diagonalizing the Hamiltonian and using the definition of the $ZZ$ interaction: $\zeta=E_{00}+E_{11}-E_{10}-E_{01}$, where $E_{ij}$ corresponds to the energy eigenvalue of state $i(j)$ of $Q_A (Q_B)$. If two states come into resonance, this definition can be non-trivial, as the identification of which state is, for example, the $|11\rangle$ state and which is a non-computational state, e.g., $|20\rangle$, becomes more difficult (\ref{sec:ZZtheory}). 

We observe excellent agreement between our experimental results and theoretical calculations of the variation in the $ZZ$ interaction, capturing both the regions where $\zeta$ changes sign smoothly and where it jumps discontinuously. For our multimode ring resonator, a quite small change in the frequency of one qubit, of the order of a few MHz for $Q_A$ in this case, results in a change in $\zeta/2\pi$ from zero to tens of MHz. This significant change is due to the large effective coupling between the two qubits, which is mediated by the high mode density. This large coupling causes a strong repulsion of the $|11\rangle$ and $|20\rangle$ states, leading to large changes in $\zeta$ (\ref{sec:ZZdetails}).

\section{Discussion} 
We have demonstrated that the metamaterial ring resonator exhibits a 
broad range of bipolar interactions between transmon qubits and the ring resonator modes, as well as interqubit interactions mediated by the ring resonator. The symmetry of the ring resonator geometry leads to even or odd parity of the $g$-coupling for each mode to the two qubits, a property that 
is unique to the ring geometry, compared to a linear left-handed transmission line configuration. If the degeneracy between the opposite parity $g$-coupled modes is broken by an asymmetry in the device layout, tuning the qubits between such a pair of modes can lead to the cancellation of $ZZ$ interactions between the qubits. 
In addition, due to the dense spectrum of the ring resonator modes and the different parities of the $g$-couplings, there is also a large variation of the $J$-coupling between qubits as their frequency is varied. Therefore, we can make a large change in both the $J$-coupling and the $ZZ$ interaction through relatively small changes in qubit frequency.

The large range of tunability for interqubit interactions suggests a potential path for future work based on this device design investigating $ZZ$-based two-qubit gates. Prior work on two-qubit gates based on in situ modulation of the $ZZ$ interaction using tunable couplers has required tuning of at least one circuit element over a frequency range of the order of GHz \cite{xu2020high,stehlik2021tunable}. Fast tuning over such a large range introduces the risk of fidelity degradation through leakage to other modes, both intentional and spurious, within this frequency window. By contrast, the orders of magnitude smaller tuning range required for moving between the on and off regimes of the $ZZ$ interaction for our multimode ring resonator provides a promising pathway for implementing a high-fidelity two-qubit entangling gate. Gate errors due to Purcell losses caused by coupling between the qubits and the multiple nearby modes could be reduced by eliminating the separate feedline coupled to the ring resonator, thus making the modes limited by internal rather than external losses.

This platform could also be used for implementing a non-$ZZ$-based two-qubit gate, for example, based on cross resonance \cite{Rigetti2010,Chow2011}. By operating the qubits at frequencies where the $ZZ$-interaction is minimized but the $J$-coupling is large, one could potentially perform a fast cross resonance entangling operation while minimizing gate errors caused by the parasitic $ZZ$ interaction that occurs in general for systems of coupled transmons \cite{Tripathi2019,magesan2020effective}. As an example, for this device, as shown in Fig.~\ref{fig:ZZ}(a), for a dressed frequency for $Q_A$ of 5.195~GHz and dressed frequency for $Q_B$ of 5.04~GHz, the $ZZ$ interaction has a zero-crossing; at these same qubit frequencies where $\zeta\slash2\pi=0$, the $J$-coupling computed from Eq.~(5) is $\sim$11~MHz. 

In addition, based on the mode structure presented above, the compact ring geometry allows for coupling to more than two qubits, with the possibility of tuning $ZZ$ interactions or $J$-couplings between different pairs of qubits with selective flux control of each qubit. Coupling a larger number of qubits around the ring, or even coupling together several ring resonators, each coupled to multiple qubits, could be a path for creating a multi-qubit system with long-range and tunable coupling between physically distant qubits. Prior work using a ring resonator architecture for long-range coupling between multiple qubits utilized a 3D ring resonator with right-handed dispersion and required more than an order of magnitude larger physical footprint \cite{hazra2021ring}. Our left-handed ring resonator geometry with its multimode spectrum can thus serve as a compact multiqubit bus for an architecture with highly tunable interactions between distant qubits.

\vspace{4mm}
\noindent\textbf{\large Acknowledgments}

\noindent This work is supported by the U.S. Government under AFOSR grant FA9550-21-1-0020. Portions of this work are supported by AFRL grant FA8750-20-1-1001. Fabrication was performed in part at the Cornell NanoScale Facility, a member of the National Nanotechnology Coordinated Infrastructure (NNCI), which is supported by the National Science Foundation (Grant NNCI-2025233). Any opinions, findings, and conclusions or recommendations expressed in this article are those of the authors and do not necessarily reflect the views of the Air Force Research Laboratory (AFRL).



\normalem
\pagestyle{fancy}
\fancyfoot[C]{ \small Approved for Public Release; Distribution Unlimited. PA \#: AFRL-2023-2427}
\fancyhead[R]{\thepage}



\newpage
\setcounter{section}{0}
\setcounter{table}{0}
\setcounter{equation}{0}
\renewcommand{\theequation}{\arabic{equation}}
\renewcommand{\thetable}{\Roman{table}}
\renewcommand*{\thesection}{Appendix \Alph{section}}

\raggedbottom

\section[Theoretical Hamiltonian and derivation of g-coupling]{Theoretical Hamiltonian\\* and derivation of {\it g}-coupling}\label{sec:gdetails}

\begin{figure}[h!]
    \centering
    \includegraphics[]{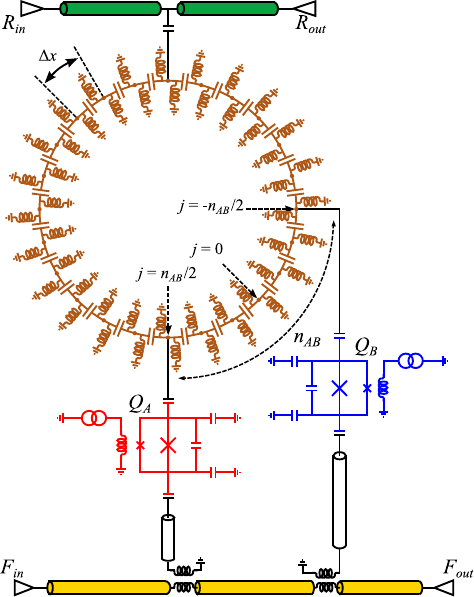}
    \caption{Full physical circuit schematic. Right-handed transmission lines labeled $R_{in}$/$R_{out}$ and $F_{in}$/$F_{out}$ in green and yellow, respectively, are used for microwave readout. The 24-cell ring-resonator circuit is shown in orange. Capacitively coupled qubits, $Q_A$ and $Q_B$, are shown in red and blue, respectively, and are separated by a distance of $n_{AB} = 6$. Qubit readout cavities are comprised of quarter-wave coplanar waveguide resonators.}
    \label{fig:circuit}
\end{figure}

\noindent We can derive an analytical expression for the $g$-coupling values of both qubits from the Lagrangian of the circuit where we ignore the effect of parasitics and therefore drop the index from $C_L$ and $L_L$. This expression is given by 
\setcounter{equation}{0}
\renewcommand{\theequation}{A\arabic{equation}}
\begin{widetext}
\begin{align}
\mathcal{L}=&\frac{1}{2}C\sum_{j=0}^{N-1} (\dot{\phi}_{j+1}-\dot{\phi}_j)^2-\frac{1}{2L}\sum_{j=0}^{N-1} \phi_j^2 +\frac{1}{2}C_{QM} (\dot{\phi}_{j_A}-\dot{\phi}_A)^2+\frac{1}{2}C_{QM} (\dot{\phi}_{j_B}-\dot{\phi}_B)^2 \nonumber\\
&+\frac{1}{2}C_{S}\dot{\phi}_A^2 +\frac{1}{2}C_{S}\dot{\phi}_B^2+E_{J}^A(\Phi_{ext}^A)\cos\left(\frac{\phi_A}{\Phi_0}\right) +E_{J}^B(\Phi_{ext}^B)\cos\left(\frac{\phi_B}{\Phi_0}\right).
\end{align}
\end{widetext}
Here, $\Phi_0$ is the magnetic flux quantum, $\phi_j$ is the flux at node $j$, $j_{A/B}$ denotes the cell number that is connected to $Q_{A/B}$, and $N$ is the total number of cells in the ring resonator. In the following, we will assume that the number of cells between the two qubits $n_{AB} \coloneqq j_A-j_B$ is even. This matches the setup of the experimental device, where $n_{AB}=6$. Since $n_{AB}$ is even, we can place cell number 0 in between the qubits, which now have the indices $j_A=\frac{n_{AB}}{2}$ and $j_B=-\frac{n_{AB}}{2}$. This choice of reference frame is without loss of generality and can be seen in  Fig.~\ref{fig:circuit}. The flux-dependent Josephson energy is given by $E^q_J(\Phi^q_{ext}) = E^q_{J0}\cos{\left(\frac{\pi\Phi^q_{ext}}{\Phi_0}\right)}\sqrt{1+d_q^2\tan^2{\left(\frac{\pi\Phi^q_{ext}}{\Phi_0}\right)}}$, where $d_q = \frac{E_{J2}^q-E_{J1}^q}{E_{J1}^q+E_{J2}^q}$ and $E_{J1}^q$ and $E_{J2}^q$ is the maximum Josephson energy of the two junctions that make up the SQUID loop. The external flux, $\Phi^q_{ext}$, gives the flux coupled to the SQUID loop for each qubit $q$.
$C_S$ is the shunt capacitance of each qubit and $C_{QM}$ is the qubit-resonator coupling capacitance.\par
We now introduce the canonical variables of Cooper pair numbers $n_j=\frac{1}{2e}Q_j=\frac{1}{2e}\frac{\partial \mathcal{L}}{\partial \dot{\phi}_j}$ and their Fourier transform $n_k=\frac{1}{\sqrt{N}}\sum_j e^{ ik\Delta xj} n_j$, where $k\Delta x/\pi\in [-1+\frac{2}{N},-1+\frac{4}{N},...,1-\frac{2}{N},1]$ form a Brillouin zone. We can immediately see that for most $k$ values there is a corresponding $-k$ mode, where positive $k$ values correspond to modes traveling around the ring resonator clockwise and negative $k$ values correspond to counterclockwise moving modes. This symmetry is essential to the device, but can be broken by the qubits, the feedline, imperfections in fabrication and wirebonds (discussion in \ref{sec:wirebond}). In our analysis, we include only the symmetry breaking due to the qubits for now, as the complexity increases with more terms. \par
The values of $k\Delta x/\pi=1$ and $k\Delta x/\pi=0$ do not have a corresponding $-k$ value, as they are at the edge and the origin of the first Brillouin zone, respectively. Note that the hypothetical state $k\Delta x/\pi=-1$ differs by $2\pi$ from $k\Delta x/\pi=1$ and is therefore not a unique state. The $k\Delta x/\pi=1$ state will later be identified as the IR mode, which is not paired with any other $-k$ state and will show no degeneracy with another mode. A similar argument can be made for the $k\Delta x/\pi=0$ state, which corresponds to the highest frequency mode. In the following we will focus on the paired states, since we are interested in an equation for $g(\omega)$. An equation for the $g$-coupling of the IR mode needs to be calculated separately, which can be done in a similar manner, but will not be shown here.\par
For the Fourier-transformed canonical variables, we get \newline
\begin{widetext}
\begin{align}
\dot{\phi}_{k} &=L\omega^2(k)\Big( 2en_{-k} + \frac{1}{\sqrt{N}}C_{QM} (\dot{\phi}_{n_{AB}/2}-\dot{\phi}_A)e^{-ik\frac{x_{AB}}{2}} +\frac{1}{\sqrt{N}}C_{QM} (\dot{\phi}_{-n_{AB}/2}-\dot{\phi}_B)e^{ik\frac{x_{AB}}{2}}\Big),
\end{align}
\end{widetext}
\noindent with the distance between the qubits given by $x_{AB}=n_{AB}\Delta x $, and the left-handed dispersion relation given by $\omega^2(k)=\frac{1}{4CL}\frac{1}{\sin^2\left(\frac{k\Delta x}{2}\right)}$. The scaling with $\omega^2$ is unique to left-handed materials and causes a different frequency dependence for the $g$ values than for a system with qubits coupled to a right-handed set of modes. Here $\dot{\phi}_k$ is expressed in terms of $n_{-k}$, further highlighting the symmetry between $k$ and $-k$, which is only broken by the qubit terms. Next, we renormalize the flux to $\varphi=\phi/\Phi_0$, ignore the coupling to the IR mode and the highest frequency mode, and only focus on the paired modes. We transform into even and odd modes using the change of variables
\begin{align*}
n_{k,E/O}=&\frac{1}{\sqrt{2}}\Big(n_k\pm n_{-k}\Big),\\
\varphi_{k,E/O}=&\frac{1}{\sqrt{2}}\Big(\varphi_k\pm \varphi_{-k}\Big),
\end{align*}
and introduce the notation $C_{rat}=\frac{C_{QM}}{C_{QM}+C_S}$. The total Hamiltonian consists of the parts $H_{tot}=H_{res}+H_{A}+H_{B}+H_{AB}^{int}+H_{mode}^{int}+H_{E}^{A}+H_{O}^{A}+H_{E}^{B}+H_{O}^{B}$. All of these terms will be defined in the remainder of this section. Without proof we here assume the interaction between resonator modes $H_{mode}^{int}$ is small compared to the other terms, which is supported by good agreement with numerical results. This approximation breaks down for modes that are very close to the IR mode. For the 24-cell ring resonator, no modes are close enough to show a noticeable deviation from theory. In the main text, we also introduce a 240-cell ring resonator. Here, some of the modes are close enough to the IR mode (less than 10~MHz away) and therefore they will show a difference in coupling strength for numerical and analytical results. For the Hamiltonian of the modes and qubits we get
\begin{align*}
H_{res}=&2e^2\sum_{k>0} L\omega^2(k)(n_{k,E}^2-n_{k,O}^2) \\
&+ \frac{\Phi_0^2}{2L}\sum_{k>0}(\varphi_{k,E}^2-\varphi_{k,O}^2),\\
H_{A/B}=&\frac{2e^2}{C_S}\Big(1-\frac{1}{1-s_{AB}^2}K^2C_{rat}^2\frac{C_S}{C_{QM}}\Big)n_{A/B}^2\\
&-E_{J}^{A/B}\cos(\varphi_{A/B}),
\end{align*}
with $\frac{1}{K}=1+\frac{C_{QM}}{N}(1-C_{rat})\sum_k L\omega^2(k)$ and $s_{AB}= K\frac{C_{QM}}{N}(1-C_{rat})\sum_k L\omega^2(k)e^{ikx_{AB}}$. The effective capacitive interaction between the two qubits mediated by the ring resonator capacitances, evaluates to

\begin{align*}
H_{AB}^{int}=&\frac{4e^2s_{AB}}{(1-s_{AB}^2)^2}\frac{K^2C_{rat}^2}{C_{QM}}n_An_B.
\end{align*}
For the interaction between a qubit and an even mode we get
\begin{align*}
H_{E}^{A/B}=&\alpha_{E}^{A/B}\frac{1}{\sqrt{N}}\sum_{k>0}\omega^2(k)\cos\Big({k\frac{x_{AB}}{2}}\Big)n_{k,E}n_{A/B}.
\end{align*}
Similarly, we can write down the interaction between a qubit and an odd mode as
\begin{align*}
H_{O}^{A/B}=\pm\alpha_{O}^{A/B}\frac{1}{\sqrt{N}}\sum_{k>0}\omega^2(k)\sin\Big({k\frac{x_{AB}}{2}}\Big)n_{k,O}n_{A/B}.
\end{align*}
Here, the $+$ branch is for qubit $A$ and $-$ is for $B$.
If we assume both qubits have the same shunt capacitance and the same coupling capacitance, the prefactors are independent of the qubit and given by
\begin{align*}
\alpha_{E}=&2\sqrt{2}e^2L\frac{K^2}{1-s_{AB}^2}\frac{1}{1+s_{AB}}C_{rat}(1-C_{rat}),\\
\alpha_{O}=-&2\sqrt{2}e^2L\frac{K^2}{1-s_{AB}^2}\frac{1}{1-s_{AB}}C_{rat}(1-C_{rat}).
\end{align*}
We can express the flux and charge operators in terms of creation and annihilation operators $a^\dagger$ and $a$:
\begin{align*}
\varphi=&\varphi_{zpf}(a+a^\dagger),\\
n=&in_{zpf}(a-a^\dagger).
\end{align*}
We can identify the zero point fluctuation of the charge as $n_{zpf}=\sqrt{\frac{\hbar}{2eL\omega}}$, which scales with $1/\sqrt{\omega}$. Similarly, the zero point fluctuations of the charge for each qubit scales with $\sqrt{\omega_{A/B}}$. We redefine the proportionality factors $\alpha_{E/O}^{A/B}$ to include the zero point fluctuations, except for their frequency dependence, and label them as ${\alpha_{E/O}^{A/B}}'$. By only considering terms that conserve excitation number based on the rotating wave approximation (RWA), the coupling Hamiltonian reduces to \newline
\begin{widetext}
\begin{align*}
H_{E}^{A/B}=&{\alpha_{E}^{A/B}}'\frac{1}{\sqrt{N}}\sum_{k>0}\sqrt{\omega_{A/B}}\omega^{\frac{3}{2}}(k)\cos\Big({k\frac{x_{AB}}{2}}\Big)(a_{k,E}^\dagger a_{A/B} + a_{k,E} a_{A/B}^\dagger),\\
H_{O}^{A/B}=\pm&{\alpha_{O}^{A/B}}'\frac{1}{\sqrt{N}}\sum_{k>0}\sqrt{\omega_{A/B}}\omega^{\frac{3}{2}}(k)\sin\Big({k\frac{x_{AB}}{2}}\Big)(a_{k,O}^\dagger a_{A/B} + a_{k,O} a_{A/B}^\dagger).
\end{align*}
\end{widetext}
\noindent With this Hamiltonian we arrive at an expression for the $g$-coupling values as a function of $k$ and $\omega(k)$
\begin{align*}
g_{E}^{A/B}(k) = &{\alpha_{E}^{A/B}}'\frac{1}{\sqrt{N}}\sqrt{\omega_{A/B}\omega^3(k)} \cos\left(k\frac{x_{AB}}{2}\right),
\end{align*}
\begin{align*}
g_{O}^{A/B}(k) = \pm&{\alpha_{O}^{A/B}}'\frac{1}{\sqrt{N}}\sqrt{\omega_{A/B}\omega^3(k)} \sin\left(k\frac{x_{AB}}{2}\right).
\end{align*}
\noindent Using the dispersion relation of left-handed materials, we can also express this in terms of the frequency. In the case of a simple dispersion relation without parasitics, this gives us
\begin{widetext}
\begin{align*}
g_{E}^{A/B}(\omega_{rE}) = &{\alpha_{E}^{A/B}}'\frac{1}{\sqrt{N}}\sqrt{\omega_{A/B}\omega_{rE}^3} \cos\left(\frac{n_{AB}}{2}\arcsin\left(\frac{\omega_{IR}}{\omega_{rE}}\right)\right),
\end{align*}
\begin{align*}
g_{O}^{A/B}(\omega_{rO}) = \pm&{\alpha_{O}^{A/B}}'\frac{1}{\sqrt{N}}\sqrt{\omega_{A/B}\omega_{rO}^3} \sin\left(\frac{n_{AB}}{2}\arcsin\left(\frac{\omega_{IR}}{\omega_{rO}}\right)\right).
\end{align*}
\end{widetext}
\noindent Note that the proportionality constants ${\alpha_{E/O}^{A}}'$ and ${\alpha_{E/O}^{B}}'$ can be either positive or negative, but always have the same sign. Therefore we do not determine the actual sign of the $g$-couplings to the modes here, only the relative sign between the interactions of each mode to $Q_A$ and $Q_B$ is given. We refer to this as the parity of the modes or the parity of the $g$-couplings. Modes with even parity share the same sign between $Q_A$ and $Q_B$ and modes with odd parity have opposite signs for $Q_A$ and $Q_B$.

\section{Device Fabrication}\label{sec:Fabrication}
\setcounter{equation}{0}
\renewcommand{\theequation}{B\arabic{equation}}
The base layer of the ring resonator device is comprised of 60-nm thick sputtered niobium on a high resistivity ($>$ 10 k$\Omega$-cm) 100-mm silicon wafer. Before sputtering, the silicon wafer undergoes a standard RCA clean process and an oxide etch in a buffered 2$\%$ HF solution to remove native oxides. The photolithography patterning is performed by a DUV wafer stepper, and the niobium is etched in an ICP etcher using BCl$_3$, Cl$_2$,
and Ar. All resist is stripped in a TMAH hot strip bath and an oxygen plasma cleaning is performed to remove residual resist residue. We then perform a second 20$\%$ buffered HF etch 
to reduce surface oxides. The device has ground straps along the feedlines labeled $R_{in}/R_{out}$ and $F_{in}/F_{out}$, as well as the flux-bias lines. Patches of evaporated SiO$_2$ isolate the straps from the signal traces, and sputtered aluminum provides the electrical ground connection, both patterned via a standard lift-off process. The junctions are patterned using 100~keV electron-beam lithography of a PMMA/MMA bilayer resist stack, then deposited via a conventional double-angle shadow-evaporation process. \\

\section{Device parameters}\label{sec:Device}
\setcounter{equation}{0}
\renewcommand{\theequation}{C\arabic{equation}}
\begin{table}[h!]
\begin{center}
\begin{tabular}{|clll|}
\hline
\multicolumn{4}{|c|}{Ring resonator cell parameters}  \\ \hline
\multicolumn{1}{|l|}{Parameter}  & \multicolumn{1}{l|}{Simulation} & \multicolumn{1}{l|}{Theory} & Description                 \\ \hline
\multicolumn{1}{|l|}{$L_L$ (nH)} & \multicolumn{1}{l|}{0.8} & \multicolumn{1}{l|}{1.04} & Unit cell inductance        \\ \hline
\multicolumn{1}{|l|}{$C_L$ (fF)} & \multicolumn{1}{l|}{303} & \multicolumn{1}{l|}{303} & Unit cell capacitiance      \\ \hline
\multicolumn{1}{|l|}{$L_R$ (nH)} & \multicolumn{1}{l|}{0.12}  & \multicolumn{1}{l|}{0.12} & Unit cell stray inductance  \\ \hline
\multicolumn{1}{|l|}{$C_R$ (fF)} & \multicolumn{1}{l|}{50}  & \multicolumn{1}{l|}{50}  & Unit cell stray capacitance \\ \hline
\multicolumn{1}{|l|}{$L_W$ (nH)} & \multicolumn{1}{l|}{1.2}   & \multicolumn{1}{l|}{1.5}  & Wirebond inductance \\ \hline
\multicolumn{1}{|l|}{$C_C$ (fF)} & \multicolumn{1}{l|}{371}    & \multicolumn{1}{l|}{371} & Center disk capacitance \\ \hline
\end{tabular}
\caption{\label{Tab:RRparams}\justifying Ring resonator cell parameters obtained via finite element simulation. Unit cell capacitance, $C_L$ and total center disk capacitance, $C_C$, obtained using Ansys Q3D software. Unit cell inductance and wirebond inductance, $L_L$ determined using InductEx software. $L_W$ is calculated as 1 nH/mm \cite{wenner2011wirebond}. The stray inductance and capacitance, $L_R$ and $C_R$, were found using Sonnet software to identify the self resonances of the lumped-element unit-cell structures. The values in the theory column come from modeling of the device and adjusting parameters to match the measured ring resonator spectrum. The discrepancy between the theoretical and simulated values for $L_L$ are likely due to imperfect grounding and approximations used in estimating the inductive contributions of the wirebonds based on extracted lengths from microscope images.}
\end{center}
\end{table}

A full circuit diagram of the device is shown in  Fig. \ref{fig:circuit}. Parameters for the left-handed ring-resonator are given in Table \ref{Tab:RRparams}. The cell capacitances and inductances were determined via finite element simulation. For all theory calculations and modeling we included the unit-cell stray inductance $L_R$ and $C_R$. The circuit diagram for a unit cell including $L_R$ and $C_R$ is shown in  Fig.~\ref{fig:wirebonds}(c).

The resultant band gap at low frequencies due to the left-handed dispersion of the resonator is bounded by the infrared (IR) cutoff frequency, which can be calculated from the cell parameters of the device. Accounting for only the unit cell capacitances and inductances, the IR cutoff for this device is simulated to be 5.6~GHz. Due to the parasitic stray reactances of the ring resonator lumped elements and the grounding wirebonds, the measured IR cutoff is substantially lower at 4.28~GHz. The theoretical infrared cutoff, including the wirebond model and stray reactances, is 4.28~GHz,  as shown in Fig.~\ref{fig:wirebonds}.

\begin{center}
\begin{table*}[]
\begin{tabular}{|p{1.5cm}|p{1.6cm}|p{6cm}|p{6.5cm}|}
\cline{1-4}
Parameter & Value & Description & Method of Determination      \\ \hline        
$E^A_{J0}$ & 23.8 GHz  & Maximum Josephson energy, $Q_A$  & Least squares minimization of Hamiltonian w.r.t. qubit spectroscopy data \\ \hline
$E^B_{J0}$ & 25.7 GHz  & Maximum Josephson energy, $Q_B$   & Least squares minimization of Hamiltonian w.r.t. qubit spectroscopy data \\ \hline
$E^A_C$  & 243 MHz   & Charging energy, $Q_A$   & Measurement of $f_{01}$ and $f_{02}/2$ transition at LSS and transmon modeling               \\ \hline
$E^B_C$  & 223 MHz   & Charging energy, $Q_B$    & Measurement of $f_{01}$ and $f_{02}/2$ transition at LSS and transmon modeling         \\ \hline
$d_A$ & 0.38  & Junction asymmetry, $Q_A$ & Least squares minimization of Hamiltonian w.r.t. qubit spectroscopy data \\ \hline
$d_B$ & 0.39  & Junction asymmetry, $Q_B$   & Least squares minimization of Hamiltonian w.r.t. qubit spectroscopy data \\ \hline
$T^A_1$  & 19.1 $\mu$s   & Relaxation time at $\Phi/\Phi_0 = 0.5$, $Q_A$    & Independent measurement                \\ \hline
$T^B_1$  & 19.4 $\mu$s   & Relaxation time at $\Phi/\Phi_0 = 0.5$, $Q_B$    & Independent measurement                \\ \hline 
$f_C^A$  & 6.358 GHz  & Bare cavity frequency, $Q_A$    & High power $S_{21}$ measurement \\ \hline
$f_C^B$  & 6.166 GHz  & Bare cavity frequency, $Q_B$    & High power $S_{21}$ measurement \\ \hline
$Q_C^A$  & 3,590  & Coupling quality factor of cavity, $Q_A$  & Inverse mapping fit of high power cavity scan \\ \hline
$Q_C^B$  & 3,290  & Coupling quality factor of cavity, $Q_B$    & Inverse mapping fit of high power cavity scan \\ \hline
$g_R^A/2\pi$  & 40 MHz  & Coupling between qubit and readout cavity, $Q_A$  & Dispersive shift measurements \\ \hline
$g_R^B/2\pi$  & 45 MHz  & Coupling between qubit and readout cavity, $Q_B$    & Dispersive shift measurements  \\ \hline
$C_S$    & 83 fF   & Shunt capacitance   & Finite element simulations   \\ \hline
$C_{QR}$   & 2.5 fF   & Qubit-readout resonator coupling capacitance & Finite element simulations    \\ \hline
$C_{QM}$  & 17.5 fF  & Qubit-ring resonator coupling capacitance    & Finite element simulations    \\ \hline
\end{tabular}
\caption{\label{Tab:Qubits} Parameters for two transmon qubits, $Q_A$ and $Q_B$.}
\end{table*}
\end{center}
Qubit details are given in Table \ref{Tab:Qubits}. The effective qubit capacitances are determined via finite element simulation and converting the floating style qubit circuit to the effective transmon circuit using the method described in Ref.~\cite{koch2007charge}. The values for $E_C^A$ and $E_C^B$ were calculated using Hamiltonian modeling of $E_C$ as a function of anharmonicity extracted from independent measurement of the qubit's $f_{01}$ and $f_{02}/2$ transition frequencies. We determined $E_{J0}^A$ and $E_{J0}^B$ by fitting spectroscopy data to a characteristic asymmetric transmon curve. This was done because we did not directly measure the qubit maximum transition frequency due to challenges in the chip design. The bare readout cavity frequency for $Q_A$ is 6.358~GHz and 6.166~GHz for $Q_B$. From fits, we calculate the maximum transition frequencies for $Q_A$ and $Q_B$ to be 6.58~GHz and 6.59~GHz, respectively. The proximity of the qubit upper sweetspot to the readout cavity and nearby mode frequencies results in  Purcell loss and broad qubit linewidths, making a direct determination of the maximum transition frequencies of the qubits challenging. \par

\twocolumngrid
\section{Wirebond model}\label{sec:wirebond}
\setcounter{equation}{0}
\renewcommand{\theequation}{D\arabic{equation}}
\begin{figure}[]
    \centering
    \includegraphics[]{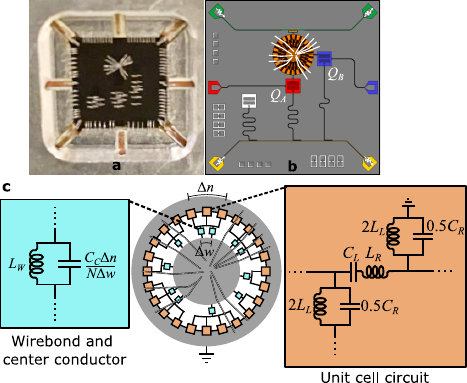}
    \caption{(a) Image of wirebonds used for grounding the device, as well as attachments to signal traces. (b) GDS of device layout including approximate wirebond positions shown with white lines. (c) Theoretical wirebond model used for modeling mode resonances and coupling strengths.}
    \label{fig:wirebonds}
\end{figure}
The degeneracy-breaking effect of utilizing wirebonds to ground the center conductor of the ring resonator results in pairs of modes above the infrared cutoff frequency. This is highlighted in Fig.~\ref{fig:chip}(e) of the main text, where we show that by including the stray capacitances and inductances for the lumped circuit elements that comprise the ring resonator cells, as well as including a model accounting for the imperfect grounding of the center disk and inductive contributions of the wirebonds, we obtain close agreement with the measured mode spectrum. This imperfect grounding of the center disk requires us to account for the shunting capacitance of the central disk of the ring resonator, which we call $C_C$, and calculate via finite element simulations, as shown in Table~\ref{Tab:RRparams}. An image of the physical device with wirebonds is shown in  Fig.~\ref{fig:wirebonds}(a), and the device GDS with the location of the wirebonds is shown in (b). Figure~\ref{fig:wirebonds}(c) shows the circuit used to theoretically model the ring resonator modes. The orange circuit shows the unit cell of a left-handed metamaterial accounting for stray reactances, resulting in a system that is left-handed up until an ultraviolet cutoff frequency equal to the minimum self resonance frequency of the left-handed circuit components. For this device, the self resonance of the inductive component $\omega_L/2\pi = 1/\sqrt{L_LC_R}$ is 25.2 GHz, below which the dispersion is left-handed and the self resonance of the capacitive component $\omega_C/2\pi = 1/\sqrt{C_LL_R}$ is 26.4 GHz, above which the dispersion becomes right-handed \cite{wang2019mode}. The wirebonds highlighted in blue account for the capacitance of the center disk and an inductance of roughly 1~nH/mm resulting in $1.5$~nH per wirebond \cite{wenner2011wirebond}. Ideally wirebonds are short, evenly spaced and densely packed. Due to the structure of our ring resonator, the wirebonds need to avoid the space that the qubits and the feedline occupy, therefore it is impossible to space the bonds evenly, and some areas will have a higher wirebond density than others. This irregular density and positioning of the wirebond connections on the center disk breaks the rotational symmetry of the ring resonator and therefore also breaks the symmetry between clockwise and counterclockwise moving waves and even and odd modes. Because we use numerical methods to simulate the circuit,  the position of the wirebond model elements does not exactly follow the experimental placement, as we are focused primarily on achieving an effective model that breaks the symmetry of the circuit with comparable strength.

\section{Cryostat and wiring setup}\label{sec:Fridgewiring}
\begin{center}
\begin{figure*}[]
    \centering
    \includegraphics[]{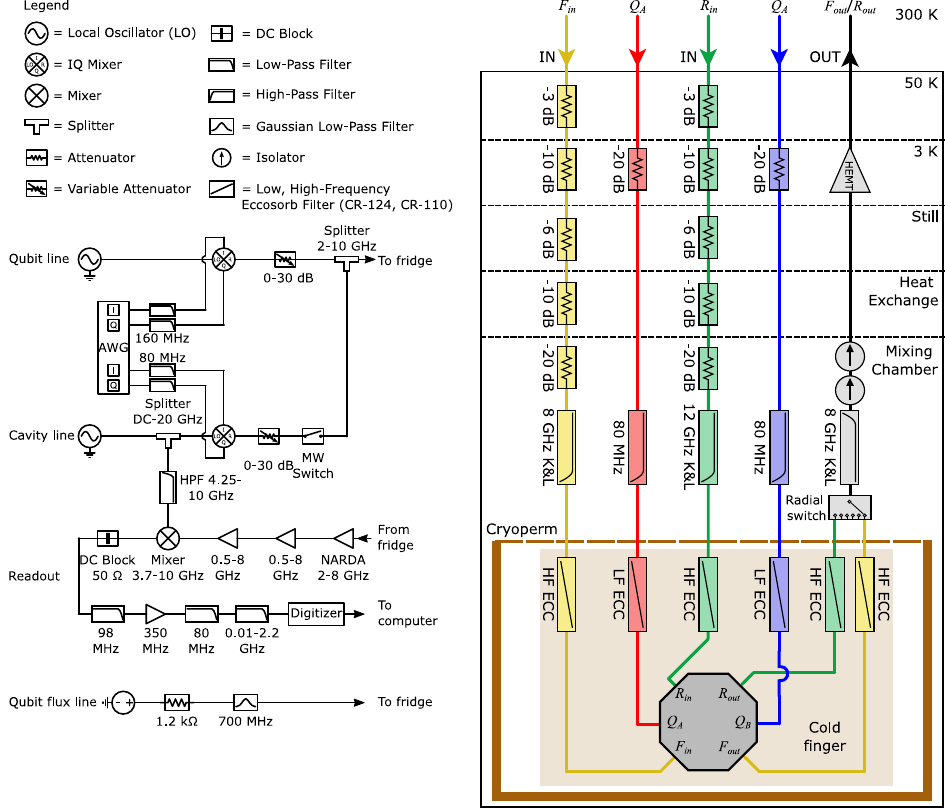}
    \caption{Wiring schematic for room temperature and cryogenic cabling down to sample box. Standard heterodyne setup is used for measurement. All filters are low-pass unless otherwise indicated.}
    \label{fig:wiring}
\end{figure*}
\end{center}

\setcounter{equation}{0}
\renewcommand{\theequation}{E\arabic{equation}}
Measurements are performed on a dilution refrigerator below 15~mK. Figure \ref{fig:wiring} shows details on the cabling and shielding setup on the dilution refrigerator, as well as the room-temperature electronic configuration for input signals and readout. An infrared-absorbent layer is applied to the inner surfaces of the Cryoperm magnetic shield and the MXC shield \cite{barends2011minimizing}. A Radiall relay switch on the output line allows switching between measurements via the two feedlines, $R_{in}/R_{out}$ and $F_{in}/F_{out}$.\\
\section{{\it g}-coupling measurements and fits}\label{sec:gexp}
\setcounter{equation}{0}
\renewcommand{\theequation}{F\arabic{equation}}
To extract $g_i^A$ and $g_i^B$ for each mode $i$, we modeled each mode as an independent harmonic oscillator coupled capacitively to $Q_A$ and $Q_B$, respectively. Using standard circuit quantization, we developed a reduced Hamiltonian for the circuit in the basis of transmon charge and resonator excitation number, given by \\
\begin{widetext}
\begin{align}
\hat{H}^q_{g,red}=&\left[\sum_{n}\left(4 E^q_{C}\left(n-n_{g}\right)^{2}|n\rangle\langle n|-\frac{E^q_{J}(\Phi^q_{ext})}{2}(|n+1\rangle\langle n|+| n\rangle\langle n+1|)\right)\right] \otimes \hat{\mathbb{I}}_{m} \\
&+ \sum_{i}\biggl[\sum_{m_{i}} \hat{\mathbb{I}}_{n} \otimes \hbar \omega_{i}\left(\left|m_{i}\right\rangle\left\langle m_{i}\right|+\frac{1}{2}\right) \notag\\
&+ \sum_{n, m_{i}} \hbar \vert g^q_{i}\vert n|n\rangle\langle n| \otimes \sqrt{m_{i}+1}\left(\left|m_{i}+1\right\rangle\left\langle m_{i}|+| m_{i}\right\rangle\left\langle m_{i}+1\right|\right) \otimes \hat{\mathbb{I}}_{m_{i \neq j}}\biggr] \notag
\end{align}
\end{widetext}
\noindent where $q$ is either $A$ or $B$ for $Q_A$ and $Q_B$ respectively. The variable $E^q_{C}$ is the transmon charging energy and $n_g$ is the transmon polarization charge. 
The operator $\hat{\mathbb{I}}_{m}$ is the product of the mode identity operators, $\hat{\mathbb{I}}_{m_i}$ is the $i$th metamaterial mode identity operator, and $\hat{\mathbb{I}}_{n}$ is the qubit charge basis identity operator. \par
A numerical minimization is implemented to fit the relevant eigenvalues of the Hamiltonian to the vacuum Rabi splitting data (main text, Figure \ref{fig:gcoupling}). We truncate the Hilbert space to include 8 charge states, $n$, and 4 number states, $m$, for the resonant modes in the Hamiltonian. For groups of modes with large $g$-coupling with respect to mode spacing, we include as many as three modes simultaneously in the fit. We do not explicitly include the energy contributions of the spectator qubit in the fits, but we do account for the dispersive $\chi$ shift of the ring-resonator modes due to the spectator qubit, which we have fixed at $\Phi_{ext}^q$ = 0. The free parameters in the fits are the $g^q_i$ values and $E^q_{J0}$. We vary $E^q_{J0}$ because the qubit maximum transition frequency is near the qubit readout cavity frequencies and cannot be observed directly, as explained in \ref{sec:Device}. All other parameters are determined via independent measurements and modeling.

We construct the variance-covariance matrix by taking numerical derivatives of the Hamiltonian with respect to the fit parameters. We compute 95\% confidence intervals to obtain the uncertainties in our fit parameters.

\begin{center}
\begin{table*}[t]
\begin{tabular}{|l|l|l|l|}
\cline{1-4}
\multicolumn{4}{|c|}{Qubit-ring resonator mode $g$-coupling values and parity}  \\ \hline              
Bare frequency (GHz) & $|g_i^A|/2\pi$ (MHz) & $|g_i^B|/2\pi$ (MHz)  & Parity  \\ \hline
4.2874 & 13.7  & 39.3  & Even  \\ \hline
4.3108 & 59.7  & 27.2  & Even  \\ \hline
4.3449 & 55.8  & 66.3  & Odd  \\ \hline
4.4749 & 3.68  & 9.68  & Even  \\ \hline
4.5531 & 0.0     & 26.6  & Even  \\ \hline
4.6206 & 35.8  & 24.0  & Odd  \\ \hline
4.7717 & 55.4  & 49.5  & Even  \\ \hline
4.9271 & 10.8  & 16.5  & Even  \\ \hline
5.1642 & 16.5  & 34.5  & Odd  \\ \hline
5.3049 & 36.6  & 43.4  & Even  \\ \hline
5.6856 & 80.5  & 63.0  & Odd  \\ \hline
5.8599 & 21.9  & 27.4  & Odd  \\ \hline
6.4317 & 43.9  & 37.3  & Even  \\ \hline
\end{tabular}
\caption{\label{Tab:Qubitparams} \justifying Magnitude of qubit-ring resonator $g$-coupling parameters and parity for $Q_A$ and $Q_B$. Here, the parity of the $g$-coupling values between the two qubits is determined by comparing measured versus perturbative theoretical calculations of the exchange coupling for $Q_A$ and $Q_B$, as described in the main text.} 
\end{table*}
\end{center}

\section{Details on {\it J}-coupling fits}\label{sec:Jdetails}
\setcounter{equation}{0}
\renewcommand{\theequation}{G\arabic{equation}}
The $J$-coupling fits were achieved by performing a least squares minimization of the spectroscopy data using a reduced Hamiltonian
\begin{align}
H_{J,eff}/\hbar =& \frac{\Tilde{\omega}_A}{2}\sigma_A^z + \frac{\Tilde{\omega}_B}{2}\sigma_B^z \nonumber\\
&+ J(\sigma_A^-\sigma_B^++\sigma_B^-\sigma_A^+) + 
\sum_i \omega_r^ia_i^{\dagger}a_i,
\end{align}
which reduces the qubits to two-state systems. The summation over modes, $i$, is reduced to include only the two nearest modes to the crossing of the bare frequencies of $Q_A$ and $Q_B$. The dressed qubit frequency for $Q_A$ is given by $\Tilde{\omega}_A = \omega_A+\sum_{i}\chi^A_i$. To measure $J$, the flux bias applied to the junction loop of $Q_B$ is fixed while we tune the transition frequency of $Q_A$. We obtain $\Tilde{\omega}_B$ from independent measurements in which we detuned $Q_A$.

The transition frequency from the ground to the first excited state of $Q_A$ is given by $\omega_A= \left(\sqrt{8E_J^AE_C^A}-E_C^A\right)/\hbar$. The amplitude of the exchange term for the two qubits, $J$, and the maximum Josephson energy for $Q_A$, $E_{J0}^A$, are the free parameters in the fit. The method for obtaining the magnitudes of the $g^A_i$ and $g^B_i$ couplings is outlined in ~\ref{sec:gexp}. All other parameters are determined via independent measurements.
We use the same method described in \ref{sec:gexp} in which we obtain numerical derivatives to compute 95\% confidence intervals in our fit parameters.

Table~\ref{Tab:Qubitparams} shows the magnitude of the $g$-couplings measured for this device. In the last column of Table~\ref{Tab:Qubitparams} the parity of the couplings to the modes is listed. These parities describe the signs of the coupling strength determined by the $J$-coupling, as explained in the main text. In the basic theoretical model, the mode at $4.927$~GHz couples with $g=0$ to both qubits and is supposed to have an odd parity. However the experimental $J$ analysis determined that the parity here is even. This can happen for values where theory predicts zero coupling, but symmetry-breaking effects push the $g$ value away from zero. These asymmetric contributions do not need to follow the parity assigned by theory, and in the case of the $4.927$~GHz mode, both modes can be pushed to positive $g$ values, even though theory predicts opposite signs. This effect is only relevant for the smallest $g$ values, since the asymmetric contributions are not strong enough for larger values.

\section[Perturbative and non-perturbative J calculations] {Perturbative and non-perturbative\\*{\it J} 
calculations}\label{sec:Jtheory}
\setcounter{equation}{0}
\renewcommand{\theequation}{H\arabic{equation}}
\begin{figure}[h!]
    \centering
    \includegraphics[]{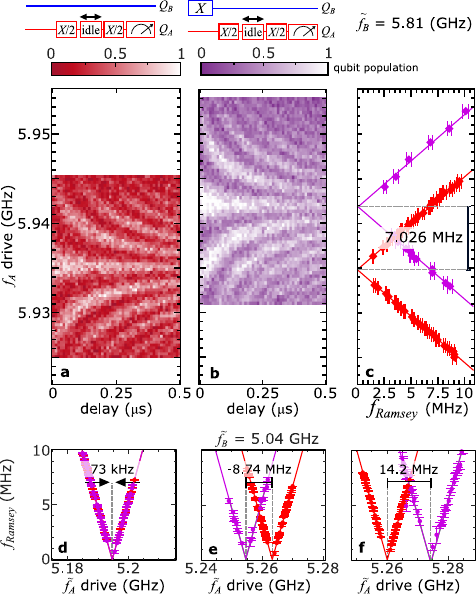}
    \caption{Measurement sequence for obtaining $ZZ$ interaction strength. (a) A Ramsey pulse sequence is performed on $Q_A$ at variable drive frequencies while $Q_B$ is idle. (b) A second Ramsey fringe measurement is performed on $Q_A$ after a $\pi$-pulse is applied to $Q_B$. (c) Horizontal slices of Ramsey fringe measurements are fit to obtain the Ramsey oscillation frequency, $f_{Ramsey}$, at each drive frequency. The $ZZ$ interaction, denoted $\zeta/2\pi$, is the change in the dressed transition frequency for $Q_A$, $\tilde{f_A}$, depending on whether $Q_B$ is in the ground or excited state. (d-f) Three example plots of Ramsey frequency as a function of $Q_A$ drive frequency ($\tilde{f_A}$ drive) for different $Q_A$ bias points with linear fit lines for $ZZ$ measurements taken when $Q_B$ is at 5.04~GHz.}
    \label{fig:ZZsup}
\end{figure}

The theoretical $J$-coupling strengths are computed using both perturbative and non-perturbative methods. The perturbative Schrieffer Wolff method
\begin{equation} \label{eq:J}
    J = \frac{1}{2}\sum_{i}g^A_ig^B_i\left(\frac{1}{\Delta^A_i} + \frac{1}{\Delta^B_i}\right)
\end{equation}
is also highlighted in the main text. Here, it is notable that $g_i^Ag_i^B$ scales with $\omega^3$, while $\Delta_i$ scales with $\omega$ if the resonator frequency is much larger than the qubit frequency. This means that high-frequency modes still contribute some effective coupling strength, even if they are far away from the qubits. Because of this, we sum over all the modes for which we have experimental $g$ values, as well as those that are above the upper sweet spot of the qubits. For those we tune the qubits to the upper sweet spot and observe a change in the dressed resonator frequency. Based on this we can extract a $g$ value for each mode, which is used to model the interaction with these high-frequency modes. \par

We want to confirm our theoretical predictions of the effective $J$-coupling also with a non-perturbative method, since this type of `long-range interaction' is not ideal for perturbative means. Therefore, we use the Least Action method \cite{Xu2021ZZfreedom,magesan2020effective,cederbaum1989block} to block-diagonalize our full-system Hamiltonian. This Hamiltonian includes single and double excitations for both qubits and resonator modes, as well as all possible combinations of one qubit excitation paired with one resonator excitation. We also include two excitations of ring resonator modes which are not the same mode. Since there are many possible combinations for this, we restrict ourselves to those that couple strongly and are relatively close to the frequency domain we are interested in. The resulting Hamiltonian is block-diagonalized using Least Action and the resulting $J_{00}$ coupling strength can be read off. This gives two $J$ values for each anticrossing between the two qubits. One is slightly before the discontinuity of the dressed frequencies and the other one is slightly after it. These two values are usually very close to each other and we average them to get one value for the anticrossing. This is then compared to Schrieffer Wolff and experimental results.

\section{{\it ZZ} measurements and fits}\label{sec:ZZdetails}
\setcounter{equation}{0}
\renewcommand{\theequation}{I\arabic{equation}}
\begin{figure}[h!]
    \centering
    \includegraphics[]{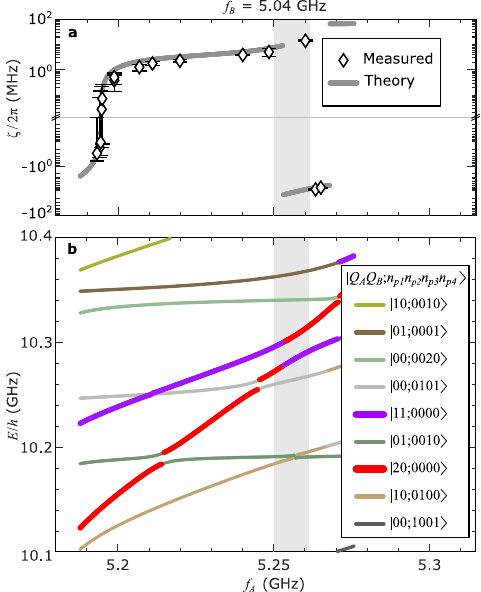}
    \caption{(a) Experimental and theoretical $ZZ$ interaction $\zeta/2\pi$ as a function of $Q_A$ frequency $f_A$ shown with diamonds and grey lines, respectively. (b) Multi-photon energies as a function of $f_A$. The light grey regions in (a) and (b) denote a region of uncertainty in the theoretical $ZZ$ calculation. The discontinuity of the $ZZ$ interaction around $5.25$~GHz is caused by an anticrossing of the $|11\rangle$ and $|20\rangle$ states sible in (b). Since there are multiple other modes close to this region, the exact point of the frequency discontinuity strongly depends on various parameters, making accurate predictions challenging.}
    \label{fig:eigen}
\end{figure}
When performing the $ZZ$ measurements, we fix the bare frequency of $Q_B$ and vary the frequency of $Q_A$, but the dressed frequency of $Q_B$ varies depending on the frequency of $Q_A$. To measure the frequency of $Q_B$ with high precision, a Ramsey fringe measurement and fit is performed for $Q_B$ for each frequency tuning of $Q_A$. The range in $\tilde{f_B}$ is 0.7~MHz (1~MHz) for the measurements in which we quote $\tilde{f_B}$ to be 5.04~GHz (5.81~GHz). These small variations in $\tilde{f_B}$ are due to the effective inter-qubit coupling via the ring resonator modes.

The pulse sequences for extracting the $ZZ$ interaction strength are shown in  Fig.~\ref{fig:ZZsup}(a,b). The Ramsey fringe fits were performed by first fitting spectroscopy data of the one-state probability for $Q_A$ as a function of idle time for each detuning frequency measured. This corresponds to taking a single horizontal slice of the data shown in  Fig. \ref{fig:ZZsup}(a,b), then applying a damped, oscillatory fit function. We remove fits with error greater than 0.4~MHz in the estimation of $f_{Ramsey}$. We then fit a line through the $f_{Ramsey}$ data with a slope of one to find the zero point for the Ramsey oscillation frequencies, as shown in  Fig.~\ref{fig:ZZsup}(c-f). We perform these fits for both pulse sequences, then calculate the difference in transition frequency for $Q_A$ to extract $\zeta/2\pi$. The error for extracting the $f_{Ramsey}$ values is computed from 95\% confidence intervals for the Ramsey fits. The error bars for the $\zeta/2\pi$ values as a function of $\tilde{f_A}$ in Fig.~\ref{fig:ZZsup} (main text) are obtained by finding the 95\% confidence intervals for a linear fit with a slope of one to the $f_{Ramsey}$ data.

\raggedbottom
\section{Theoretical {\it ZZ} calculations}\label{sec:ZZtheory}
\setcounter{equation}{0}
\renewcommand{\theequation}{J\arabic{equation}}
The $ZZ$ interaction is obtained by fully diagonalizing the Hamiltonian that was also used for calculating the $J$-coupling strengths. Alternatively, the already block-diagonalized Hamiltonian can be used to only diagonalize the qubit subspace, thus saving computational cost with similar results. The $ZZ$ interaction is then calculated using $\zeta=E_{00}+E_{11}-E_{10}-E_{01}$. The biggest impact on the $ZZ$ strength comes from the $E_{11}$ level interacting with non-computational states such as $E_{20}$. The high resonator mode density around $4-5$~GHz causes the $10$~GHz area to be highly populated with double excitations of qubits and resonator modes. For the frequency domain shown in  Fig.~\ref{fig:eigen}, the most important states are $|20;0000\rangle$ and $|01;0001\rangle$, which both couple strongly to $|11;0000\rangle$, since only one exchange of excitation is required for a transition. The anticrossing with $|20;0000\rangle$ is especially interesting here, since it causes a large shift and discontinuity of $E_{11}$, which is also observable in the experimental data in  Fig.~\ref{fig:eigen}(a). However, the experimental drop-off is positioned at $5.26$~GHz, while the theoretical one is at $5.253$~GHz. This slight disagreement is probably caused by small inaccuracies in multiple parameters. Most noticeably the anharmonicity of $Q_A$ determines the energy of the $|20;0000\rangle$ state; an inaccuracy here will result in a frequency shift of the drop-off position. There are also multiple other modes nearby, and a change of their frequencies or coupling strengths will impact the position of the discontinuity.

\normalem
\bibliographystyle{apsrev4-1}
\bibliography{bibliography}

\end{document}